\documentclass{article}

\usepackage{arxiv}
\usepackage{amsmath}
\usepackage[utf8]{inputenc} 
\usepackage[T1]{fontenc}    
\usepackage{url}            
\usepackage{booktabs}       
\usepackage{amsfonts}       
\usepackage{nicefrac}       
\usepackage{microtype}      
\usepackage{lipsum}		
\usepackage{graphicx}
\usepackage{natbib}
\usepackage{doi}
\usepackage{hyperref}       
\usepackage{fancyhdr}

\title{Escaping Geological Generative Prior Distributions in Bayesian Travel Time Tomography}

\date{} 					

\author{\hspace{1mm}Pratyay Roy \\
	School of GeoSciences\\
	University of Edinburgh\\
	Edinburgh, UK \\
	\texttt{pratyay.roy@ed.ac.uk} \\
	\And
	\hspace{1mm}Xuebin Zhao \\
	School of GeoSciences\\
	University of Edinburgh\\
	Edinburgh, UK \\
	\texttt{Xuebin.Zhao@ed.ac.uk} \\
    \And
	\hspace{1mm}Andrew Curtis \\
	School of GeoSciences\\
	University of Edinburgh\\
	Edinburgh, UK \\
	\texttt{Andrew.Curtis@ed.ac.uk} \\
}

\renewcommand{\headeright}{}

\fancypagestyle{firstpage}{%
  \fancyhf{}
  
  \fancyfoot[L]{\footnotesize Published as a conference paper at 87th EAGE Annual Conference \& Exhibition 2026 \path{https://doi.org/10.3997/2214-4609.202610891}}
  \fancyfoot[R]{\footnotesize\thepage}
}

\hypersetup{
pdftitle={Escaping Geological Generative Prior Distributions in Bayesian Travel Time Tomography},
pdfsubject={physics.geo-ph, cs.LG},
pdfauthor={Pratyay Roy, Xuebin Zhao, Andrew Curtis},
pdfkeywords={Bayesian inference, Inverse theory,  Machine learning, Monte Carlo Methods, Probability distributions,  Seismic tomography},
}

\begin{document}
\maketitle
\thispagestyle{firstpage}

\begin{abstract}
Solutions to geophysical inverse problems are generally non-unique, and greatly benefit from additional geological prior information. Generative Neural Networks (GNNs) are commonly used to parametrise such information, typically restricting solutions to a lower-dimensional latent space. The final inversion result is obtained by mapping the latent space parameters to the original higher-dimensional model space using the GNNs. However, solutions then remain confined to a low-dimensional geological generative manifold; if the true Earth model lies outside of this manifold, standard latent space inversion fails. We propose a Bayesian model space inversion method that uses the generative manifold as a guide to find the solutions that lie outside this manifold. We validate our approach in non-linear travel-time tomography experiments on both simple and more geological structures. By comparing our method with model space inversion using uniform prior probabilities and standard latent-space inversion, we demonstrate that the less informative uniform prior fails to constrain the solution, while the latent space approach fails when the true Earth model lies outside of the generative manifold. Conversely, our new model-space inversion method successfully recovers complex off-manifold geological features while maintaining low solution uncertainty.
\end{abstract}

\keywords{Bayesian inference, Inverse theory,  Machine learning, Monte Carlo Methods, Probability distributions,  Seismic tomography}

\section{Introduction}
Geophysical inverse problems almost always have non-unique solutions. To reduce non-uniqueness, geological prior knowledge is often introduced to constrain geophysical inversions. Generative Neural Networks (GNNs) can be trained to parametrise realistic, geologically plausible structures, and these networks have been used as prior information. However, published methods typically perform geophysical inversion within a low-dimensional latent space defined by the GNNs \citep{mosser_stochastic_2018, levy_variational_2023, meles_bayesian_2023, bloem_introducing_2023, santos_refining_2024}. This approach requires that the true Earth model lies within the family of models that the GNN can generate, and is therefore confined to a lower-dimensional subspace or manifold of naturally-occurring geological variability. Yet the true Earth has complexity and heterogeneity at every scale. It is therefore inevitable that the true Earth lies outside of the manifold defined by the GNNs, meaning that the prior information is known to be incorrect. To improve estimates of the solution we therefore need to move outside of the geological manifold.

We investigate a method to allow solutions that lie outside of the generative manifold to be found, but which still uses the manifold to guide the inversion. We demonstrate our method in non-linear travel-time tomography experiments, and compare the inversion results with those from the commonly used latent space inversion methods. While both methods provide reasonable inversion results if the true model solution lies within the manifold of the GNNs, our method provides far better results if the true solution lies outside of the GNN manifold.

\section{Methodology}

In a Bayesian framework, the solution to an inverse problem is described by the \textit{posterior} probability density function (pdf), \(p(\mathbf{m}| \mathbf{d}_{\rm obs})\), which is obtained by combining a \textit{prior} distribution \(p(\mathbf{m})\) that describes information known before the inversion, with the data \textit{likelihood} \(p(\mathbf{d}_{\rm obs}| \mathbf{m})\), which describes new information from the current data set. Combining these yields the posterior pdf, a relationship formally expressed by Bayes' theorem:
\begin{equation}
p(\mathbf{m}| \mathbf{d}_{\rm obs}) = \frac{p(\mathbf{d}_{\rm obs}| \mathbf{m})\,p(\mathbf{m})}{p(\mathbf{d}_{\rm obs})}
\label{eq:bayesian}
\end{equation}
The term $p(\mathbf{d}_{\mathrm{obs}})$, known as the \textit{evidence}, acts as a normalisation constant for the posterior pdf. We can formulate the posterior pdf in model space as,
\begin{equation}
   p(\mathbf{m}| \mathbf{d}_{\rm obs}) \propto \exp \left(-\frac{\|\mathbf{d}_{\text{obs}} - \textbf{f}(\mathbf{m})\|_2^2}{2\sigma_d^2}\right) \:  p(\mathbf{m})
\label{eq:model_space_bayes}
\end{equation}
where the likelihood term (first term in right hand side) quantifies the misfit between the observed data $\mathbf{d}_{obs}$ and the synthetic data, predicted by a forward operator $\textbf{f}$, given model parameter values $\mathbf{m}$. Our prior assumptions are encapsulated in the probability density function $p(\mathbf{m})$. This term will incorporate geological prior information.

Consider a GNN that defines a mapping through a decoder or generator network $\mathbf{D}: \mathbf{z} \to \mathbf{m}$, from latent parameter $\mathbf{z}$, to model parameter $\mathbf{m}$, and which has been trained such that the latent parameter approximately follow a standard normal distribution. This formulation allows us to perform Bayesian inversion directly in the latent space, estimating the posterior pdf of the latent parameter $\mathbf{z}$ by combining a data-likelihood with a standard normal prior $\mathcal{N}(0, \mathbf{I})$ on latent parameters:
\begin{equation}
p(\mathbf{z} | \mathbf{d}_{\text{obs}}) \propto  \exp\left( - \frac{\|\mathbf{d}_{obs} - \mathbf{f}(\mathbf{D}(\mathbf{z}))\|^2_2}{2\sigma_d^2} \right) \: \exp\left( - \frac{\|\mathbf{z}\|^2_2}{2} \right)
\label{eq:latent_space_bayes}\end{equation}
where $\textbf{f}(\mathbf{D}(\mathbf{z}))$ represents the forward operator acting on the decoded latent parameter. We can draw samples from the posterior $p(\mathbf{z} | \mathbf{d}_{\text{obs}})$ using Markov Chain Monte Carlo (MCMC) methods and map the results back to the high-dimensional model space through $\mathbf{\tilde{m}} = \mathbf{D}(\mathbf{z})$ to obtain the final inversion solution, $\mathbf{\tilde{m}}$. However, this approach introduces a critical limitation: if the true Earth model lies outside of the family of models the GNN (the decoder $\mathbf{D}$) can produce, even the entire distribution of the reconstructed models $\mathbf{\tilde{m}}$ will fail to represent the true solution. This occurs because GNNs capture only a finite-dimensional subset of the Earth's infinitely complex heterogeneity, whether due to imperfect prior information, inevitable training errors, and the finite number of latent parameters. As a result, the GNN produced geological models are confined to a low-dimensional subspace or manifold \citep{tenenbaum_global_2000}, known as the geological generative manifold $\mathcal{M}_{gen}$.

Ideally, if trained on geologically realistic subsurface images, the learned $\mathcal{M}_{gen}$ captures the dominant, perhaps large-scale structures (e.g., the general patterns of folded sediments or depositional sequences). The true Earth model invariably possesses complex, fine-scale geological details (e.g., thin sand lenses, or localised mineral inclusions) that lie outside of $\mathcal{M}_{gen}$. Therefore, to recover the true Earth model, it is essential to extend the search for the solution beyond the strict boundaries of $\mathcal{M}_{gen}$.

To extend the solution space beyond $\mathcal{M}_{gen}$, while still retaining it as a source of information, we introduce a prior pdf that accounts for both on-manifold features and off-manifold variability. We define this as a Gaussian probability density function $p_{gen}(\bf{m})$ over the high-dimensional model parameter space:
\begin{equation}
p_{gen}(\mathbf{m}) \propto \int_{\mathbf{y}\in\mathcal{M}_{gen}} \exp\left(-\frac{\|\mathbf{m}-\mathbf{y}\|_2^2}{2\tau^2}\right) \, d\mathbf{y}, \quad \text{where } \mathbf{y} = \mathbf{D}(\mathbf{E}(\mathbf{m})) \:,
\end{equation}
where $\tau^2$ quantifies the uncertainty of the off-manifold structures, determined from prior knowledge, trial and error on test problems, or hierarchical inversion. Here, $\mathbf{y}$ is a point on the generative manifold $\mathcal{M}_{gen}$, for a given model parameter $\mathbf{m}$ and is defined by a GNN as $\mathbf{D}(\mathbf{E}(\mathbf{m})) = \mathbf{y}$, where $\mathbf{E}$ and $\mathbf{D}$ represent an encoder and a decoder of the GNN, respectively. By solving for $\mathbf{m}$ directly in the high-dimensional model space, we effectively relax the strict $\mathcal{M}_{gen}$ constraint. While $\mathbf{y}$ (the samples from $\mathcal{M}_{gen}$) enforces geologically reasonable low-dimensional structures, the solution $\mathbf{m}$ is free to deviate from $\mathbf{y}$ to better fit the observed data $\mathbf{d}_{\text{obs}}$, thereby accounting for the complex, high-dimensional heterogeneities of the true Earth model captured in the observed data. To guide the inversion, the overall prior $p(\mathbf{m})$, is defined as the product of uniform physical bounds on parameter values $p_0(\mathbf{m})$, and our GNN-informed prior $p_{gen}(\mathbf{m})$:
\begin{equation}
p(\mathbf{m}) = p_0(\mathbf{m}) \: \int_{\mathbf{y}\in\mathcal{M}_{gen}} \exp\left(-\frac{\|\mathbf{m}-\mathbf{y}\|_2^2}{2\tau^2}\right) \, d\mathbf{y}
\label{eq:prior}
\end{equation}
By combining data-likelihood with this prior pdf, the posterior pdf can be formulated using equation \ref{eq:model_space_bayes}. MCMC methods can be used to draw samples from the posterior pdf in high-dimensional model parameter space.

\section{Travel-Time Tomography}

\begin{figure}[htbp]
    \centering
    \includegraphics[width=0.75\linewidth]{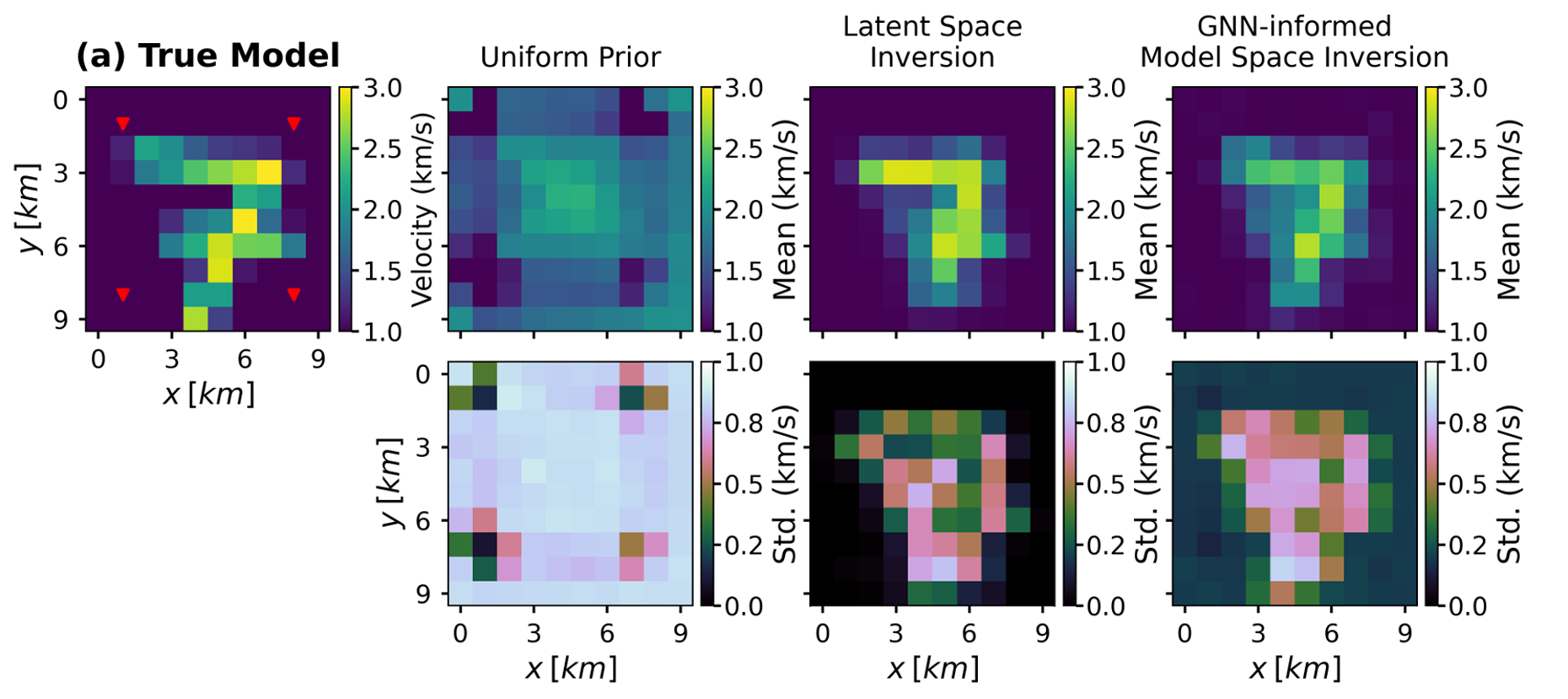}
    
    \vspace{0.5cm} 
    
    \includegraphics[width=0.75\linewidth]{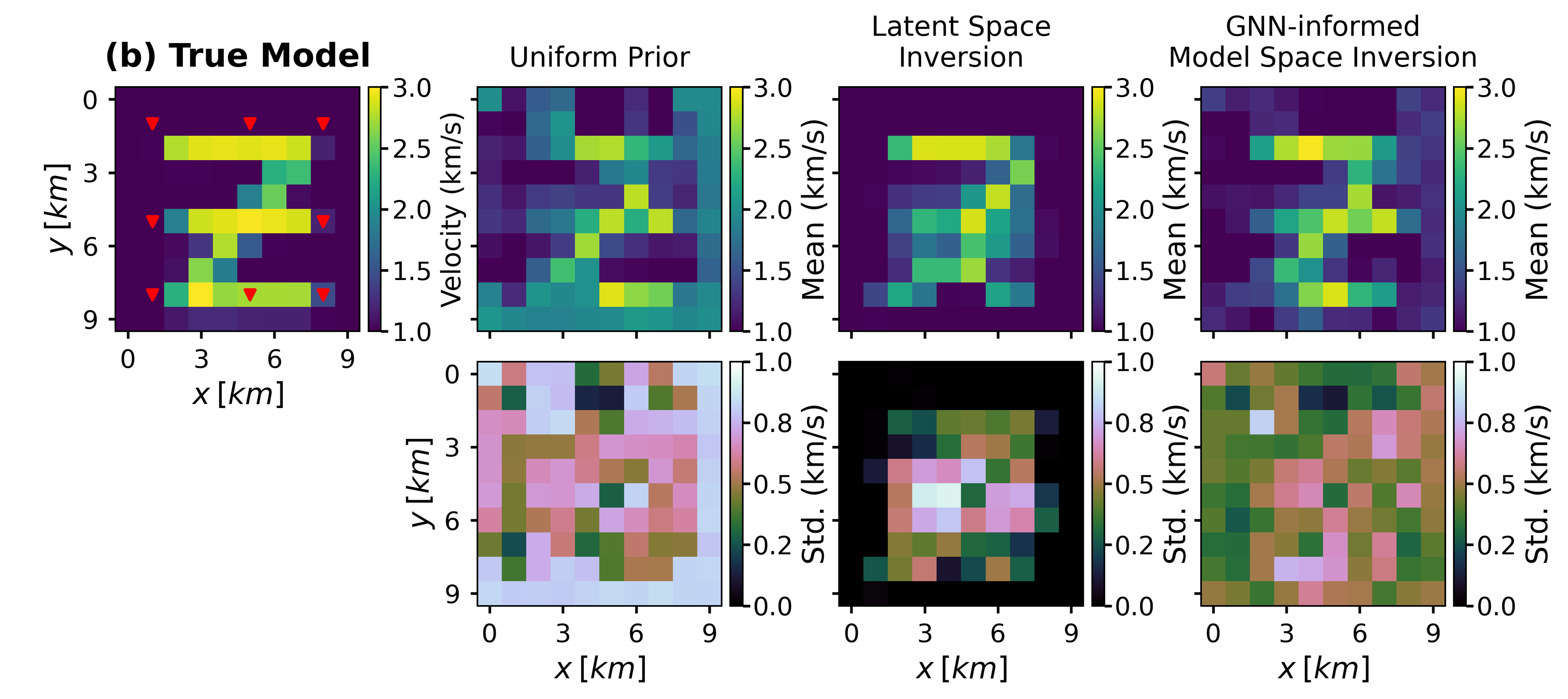}
    
    \caption{Inversion results for two simple Earth models using travel-time data between (a) 4 and (b) 8 sources/receivers (red triangles). The first column shows the true model. Subsequent columns compare Uniform Prior, Latent Space and GNN-informed Inversion.}
    \label{fig:tomography_exp_1}
\end{figure}

\begin{figure}[ht!]
    \centering
    \includegraphics[width=0.7\linewidth]{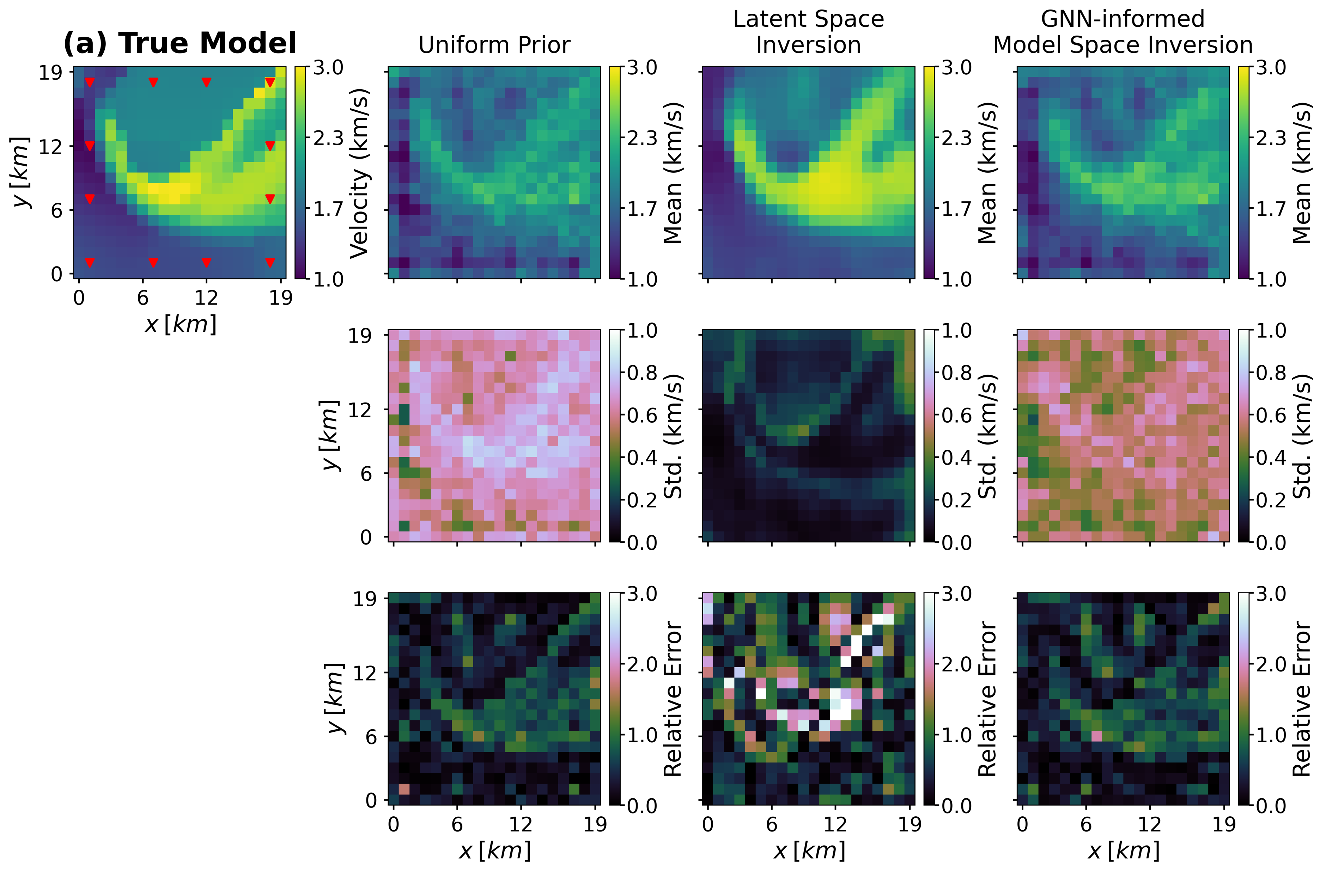}
    
    \vspace{0.5cm} 
    
    \includegraphics[width=0.7\linewidth]{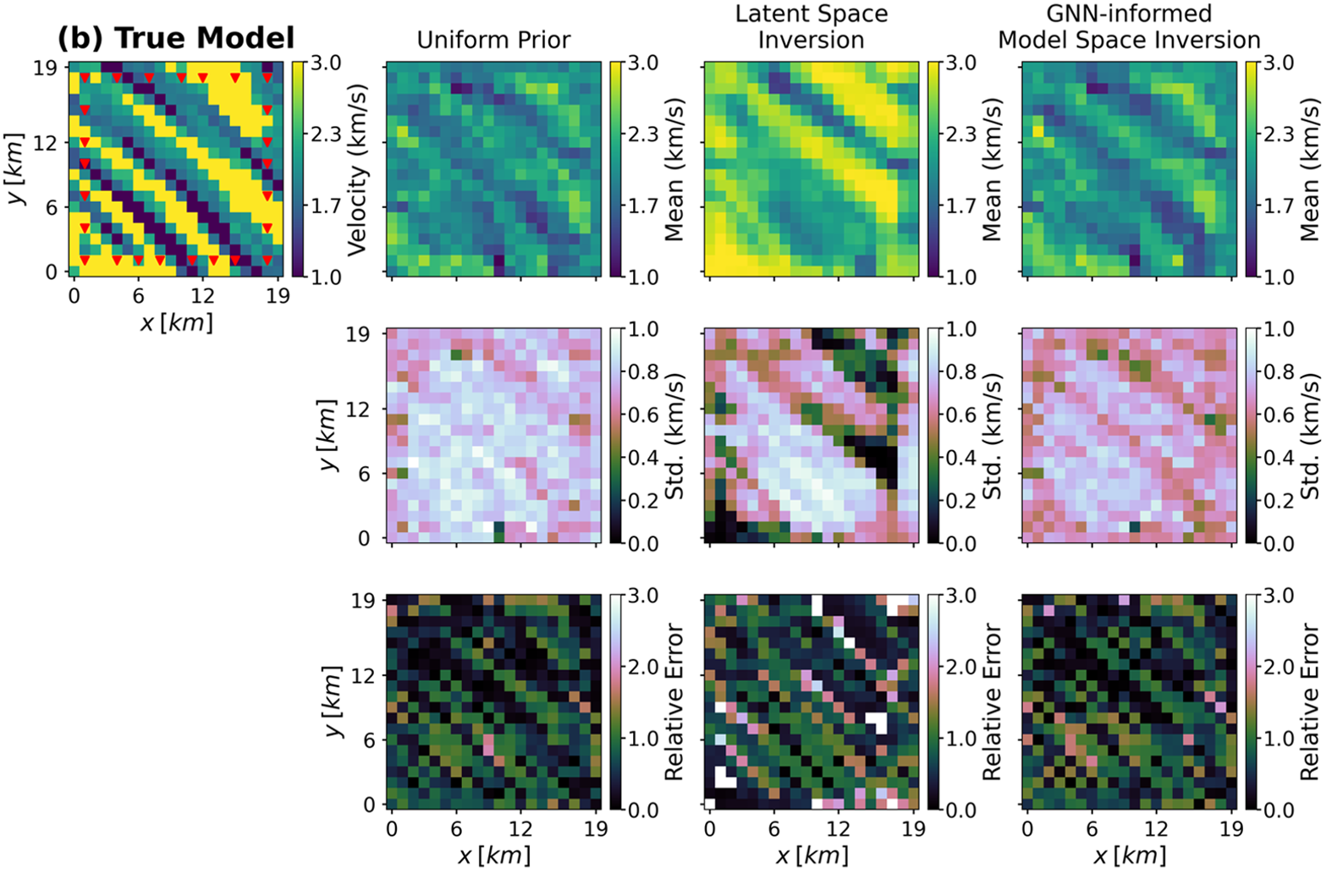}
    
    \caption{Inversion results for two different geological velocity structures using travel-time data between (a) 12 and (b) 25 sources/receivers (red triangles). True model is in the left-most columns and subsequent columns compare Uniform Prior, Latent Space and GNN-informed Inversion.}
    \label{fig:real_tomography_exp}
\end{figure}

We validate our method using four different non-linear travel-time tomography experiments. Figures \ref{fig:tomography_exp_1} and \ref{fig:real_tomography_exp} show the true velocity models, which correspond to two simple structures ($10 \times 10$ model parameters) and two more geological velocity structures ($20 \times 20$ model parameters), respectively, and all of the results. We test three different inversion strategies: to provide a baseline solution with little prior information we perform model space inversion using a uniform prior pdf (i.e., equation \ref{eq:model_space_bayes} with $p(\mathbf{m})=p_0(\mathbf{m})$), standard latent space inversion (equation \ref{eq:latent_space_bayes}), and our proposed GNN-informed model space inversion (combining equations \ref{eq:model_space_bayes} and \ref{eq:prior}). To draw samples from the posterior pdf in each example, we use the Metropolis-Hastings algorithm \citep{metropolis_equation_1953}, running a total of 40 independent chains in parallel, each with 2 million samples.

In the first tomography experiment (Figure \ref{fig:tomography_exp_1}a), we employ a sparse geometry with 4 sources and receivers, and use six independent travel-time measurements to estimate the complex velocity structure using inversion. Since the true model resembles the structure of the digit `7', we trained an Adversarial Autoencoder \citep{makhzani_adversarial_2016} on the MNIST dataset \citep{lecun_mnist_1998} (containing only the handwritten digit `7') to parametrise the prior information. This allows the network to learn the characteristic structural features of seven and incorporate them as prior information within our inversion framework. We note that while we employ an Adversarial Autoencoder here, our approach is general and applicable to any GNN. Additionally, we use the decoder of this trained Adversarial Autoencoder for the latent space inversion. By comparing each set of results (Figure \ref{fig:tomography_exp_1}a) with the true velocity model, we can observe that both latent space inversion and our GNN-informed model space inversion were able to recover the structure of the true model with very low uncertainty despite the sparse geometry. The latent space inversion (Figure \ref{fig:tomography_exp_1}a) succeeded because the true model lies within the GNN's generative manifold. Similarly, our GNN-informed model space inversion benefited from the highly informative prior information that captures the structural shape of the digit `7'. In contrast, the model space inversion with uniform prior shows a blurry posterior mean velocity plot and uniformly high posterior uncertainty.

Figure \ref{fig:tomography_exp_1}b shows the second tomography experiment, where attempt to recover a model that looks like the letter `Z'. We use the same Adversarial Autoencoder, which was trained exclusively on the digit `7', and we increase the number of sources and receivers to provide a little more information to the inversion with uniform prior information. Since the resulting generative manifold contains only `7'-like structures, our primary objective was to determine if latent space inversion could recover the off-manifold features of the true model, and to compare its performance against other inversion strategies. In Figure \ref{fig:tomography_exp_1}b, even with additional sources and receivers, the uniform prior inversion could not recover an interpretable `Z'-like structure. Likewise, because the latent space inversion result remained confined to the generative manifold, it failed to capture the true model despite its low uncertainty. Conversely, our GNN-informed approach successfully recovered the off-manifold structural features of the true model, and also achieved lower uncertainty than the inversion with uniform prior.

For more geological velocity structures (Figure \ref{fig:real_tomography_exp}a, b), we parametrise the geological prior information by training an Adversarial Autoencoder on images derived from sedimentary process model simulations. To generate the training images, we simulated 3D geological processes of deposition and erosion using SedSimple \citep{tetzlaff_stratigraphic_2023}. These 3D geological models are then converted into 3D velocity models using rock physics relations; finally, we extract horizontal cross-sections to train our GNN. 

The first case of true models containing geological structures is generated using SedSimple, but is not included in the GNN training data. We employ a sparse geometry with 12 sources and receivers (Figure \ref{fig:real_tomography_exp}a) because, if the true model falls within the generative manifold defined by the GNN then it should be able to be recovered by both GNN-informed model space inversion and latent space inversion due to the strong geological prior information. Conversely, in the second case (Figure \ref{fig:real_tomography_exp}b), the true model is not generated by SedSimple, so definitely lies outside of the GNN manifold. Since data is the only way to guide inversions beyond the manifold, we increase the number of sources and receivers to 25.

As shown in Figure \ref{fig:real_tomography_exp}a, the model space inversion with a uniform prior yields a blurry reconstruction with high uncertainty. The latent space inversion fails to capture some specific features of the true model; instead, it produces a smooth approximation of the large-scale structures that exhibits high relative error (absolute difference between true and mean velocities divided by the standard deviation at each point) in some parts, due to its misleadingly low uncertainties. By contrast, our GNN-informed model space inversion successfully recovers a posterior result with low uncertainty and low relative error. With the increased data coverage for the complex non-SedSimple in Figure \ref{fig:real_tomography_exp}b, the uniform prior inversion provides an improved mean estimate but continues to suffer from high uncertainty. The latent space inversion produces only large-scale structures because the true features lie outside the range of generative manifold (and the training set). This results in an overconfident solution with high relative error in some parts. In contrast, our GNN-informed model space inversion recovers a result with small-scale structural features, while maintaining low relative error and significantly lower uncertainty than the inversion with uniform prior.

\section{Conclusion}

In this work, we demonstrate an inversion method that incorporates geological information parametrised by the manifold of generative neural networks as prior information within a Bayesian model space inversion, but which is not absolutely constrained by that information. While non-informative uniform prior distributions perform poorly with sparse data, and latent space inversions fail when the true solution deviates from the training distribution, our proposed method allows the solution to deviate from the generative manifold, by recovering off-manifold geological features with high accuracy and low uncertainty.

\section{Acknowledgements}

We gratefully acknowledge the support provided by the Edinburgh Imaging Project sponsors, BP and TotalEnergies, for this research.

\bibliographystyle{unsrtnat}
\bibliography{eage_arxiv_ref}

\end{document}